\documentclass[%
 reprint,
 amsmath,amssymb,
 aps,
]{revtex4-2}
\usepackage[T1]{fontenc}
\usepackage{graphicx}
\usepackage{dcolumn}
\usepackage{bm}
\usepackage{booktabs} 

\begin{document}

\preprint{APS/123-QED}

\title{Remote characterization of aerogel foam concrete using dynamic speckle pattern analysis}
\author{Ramin Jamali}
\affiliation{Department of Physics, Institute for Advanced Studies in Basic Sciences (IASBS), Zanjan 45137-66731, Iran\vspace{0.1mm}}
\author{Mohammad Hadi Sadri}%
\affiliation{Department of Physics, Institute for Advanced Studies in Basic Sciences (IASBS), Zanjan 45137-66731, Iran}
\author{Ali-Reza Moradi}%
\email{moradika@iasbs.ac.ir}
\affiliation{Department of Physics, Institute for Advanced Studies in Basic Sciences (IASBS), Zanjan 45137-66731, Iran}
\affiliation{School of Nano Science, Institute for Research in Fundamental Sciences (IPM), Tehran 19395-5531, Iran}

\begin{abstract}
Aerogel foam concrete (AFC) has garnered significant attention in recent years due to its exceptional thermal insulation, lightweight structure, and versatility in construction applications. However, the durability of this material in various chemical environments, particularly acidic and alkaline solutions, remains a critical concern for its long-term performance. In this study, we employed an advanced remote characterization technique—dynamic speckle pattern analysis—to monitor and quantify the degradation and corrosion processes of AFC under these conditions. Using this non-invasive method, we extracted valuable statistical parameters, including the time history of speckle patterns, co-occurrence matrix, inertia moment, Pearson correlation, and roughness indices, to provide a comprehensive analysis of surface and structural changes. Our findings reveal that AFC exposed to acidic environments undergoes faster degradation and more severe surface damage compared to alkaline environments, as demonstrated through \textit{in situ} and remote characterization. These results underscore the importance of understanding material behavior in diverse conditions, offering critical insights for improving the durability of AFC in various applications.
\end{abstract}

\maketitle


\section{Introduction}

Aerogel foam concrete (AFC) represents a novel composite material in construction materials, which combines the ultra-lightweight and highly insulating properties of aerogels with the versatile and durable characteristics of foam concrete \cite{muralitharan2017development}. Aerogels, particularly silica-based aerogels, are among the lightest known solid materials, boasting exceptional thermal and acoustic insulation because of their porous structure and play a key role in modern architecture \cite{real2016thermal,zhang2015use,zhang2015mechanical,hu2016research}. With growing global concerns about energy consumption in buildings—especially since nearly 40\% of the total energy of the world is used in this sector—enhancing the insulating performance of construction materials has become a top priority \cite{ibrahim2015building,perez2008review}. Foam concrete, known for its low density and ease of transport, is especially useful for energy-efficient buildings but often requires modifications to improve its thermal conductivity and mechanical strength \cite{muralitharan2017development,mikulica2017foam}.
Aerogels, particularly silica aerogels, have emerged as an innovative solution to improve the insulating properties of foam concrete. Aerogels are porous materials that consist mainly of air, up to 99\% - trapped within a silica network. This unique structure gives aerogels extremely low thermal conductivity (approximately 0.015 W$/$m $\cdot$ K) and ultralow density (as low as 100 kg/m$^{3}$), making them ideal for insulation applications \cite{aegerter2011aerogels,buratti2012glazing,li2019preparation}. When incorporated into foam concrete, aerogels enhance the thermal and acoustic insulation properties without significantly compromising the lightweight nature of the material. This composite material, AFC, offers a balance of thermal performance and structural stability, making it highly desirable for energy-efficient building applications  
 \cite{li2019preparation,ng2015experimental}.

The development and application of AFC respond directly to the increasing demand for more energy-efficient and sustainable building materials. Materials such as AFCs have the potential to drastically reduce energy consumption \cite{gomes2018thermal}. Conventional building materials, such as expanded polystyrene while effective insulators, often require thick layers to achieve the necessary thermal resistance, reducing usable building space and increasing construction costs \cite{abbas2019silica,melițua2024silica}. However, AFC provides a high-performance alternative that offers superior insulation with thinner material layers, and making AFC an attractive material for modern architectural applications  \cite{muralitharan2017development,li2019preparation}. However, for AFC to be widely adopted in construction, it is essential to understand its behavior in various environmental conditions, particularly its chemical durability when exposed to aggressive agents such as acids and bases. Concrete, including AFC, is known to be susceptible to chemical degradation when exposed to acidic or alkaline environments. Acidic solutions can attack the calcium hydroxide (Ca(OH)$_{2}$) in the concrete matrix, leading to the formation of soluble salts and the leaching of material, which weakens the structure over time \cite{zivica2001acidic}. Similarly, alkaline solutions can cause alkali-silica reactions, particularly in the presence of silica-based materials like aerogels \cite{swamy1992alkali}. These reactions lead to the formation of expansive gels that can cause internal pressure, cracking, and eventual failure of the concrete structure. Therefore, it is crucial to investigate how AFC responds to chemical exposure in both acidic and alkaline environments, as this will directly impact its long-term performance and suitability for various construction applications.

To comprehensively evaluate the properties and performance of AFC, it is essential to employ robust and practical methods for characterizing both its biological and physical attributes. Common techniques used in aerogel cement characterization include thermal analysis, spectroscopy, chromatography, rheology, mechanical testing, and microscopy. Each of these methods, however, has inherent limitations \cite{li2024preparation,liu2022microstructure,gao2014aerogel}.
For instance, while scanning electron microscopy provides high-resolution 3D imaging, it is costly, offers limited scanning coverage, and necessitates an environment free from mechanical vibrations and electromagnetic interference \cite{liu2022microstructure}. Similarly, scanning force microscopy achieves nanoscale resolution and operates effectively in liquid environments, but its slow scanning rate and the potential for tip-induced specimen damage make it less suitable for observing dynamic processes \cite{perdigao2000microscopy,czepkowshi1996some}.
Reflective digital holography has emerged as a non-invasive method for real-time 3D imaging of the AFC. While this technique offers unique advantages for monitoring processes, it is highly sensitive to environmental disturbances such as vibrations and requires complex computational reconstruction. These limitations can pose challenges when applying the method to dynamic phenomena or large-scale systems \cite{celik2024imaging,she2024cement,jamali2023digital,kim2010principles,rad2021digital,jamali2023measurement,erfanifamfringe,panahi20213d}.
To assess the long-term durability of AFC, particularly in corrosive environments, non-destructive testing techniques are increasingly being used. One such technique is dynamics speckle pattern analysis (DSPA), DSPA is a remote and non-destructive optical method that allows for remote monitoring of changes and dynamics in materials \cite{goodman2007speckle,braga2003assessment}. The technique involves illuminating the sample with coherent light, typically from a laser, which produces speckle patterns—random distributions of high-contrast bright spots surrounded by dark regions. These patterns result from the interference of coherent light waves scattered by the sample \cite{goodman2007speckle,goodman1976some}. By analyzing the intensity of the speckle pattern over time, where each point contains detailed information about the material’s structure, any changes in the internal or external composition—such as those caused by chemical corrosion—are reflected in alterations in the speckle pattern \cite{aizu1991bio,fujisawa2009temperature}.

Speckle pattern formation and analysis can be achieved by using two configurations: subjective and objective. When a speckle pattern is observed and recorded at a distance from a surface that scatters light diffusely, it is referred to as ``objective.'' In this case, objective speckle patterns arise when a diffusive object is illuminated by coherent light. The size of the speckle pattern grains is influenced by the interference of waves scattered from different points on the surface. As the distance between the observation plane and the object increases, the grain size increases proportionally. These objective speckle patterns are typically observed in the far-field region, where their properties are determined by the dimensions and shape of the area illuminated by the laser beam. In contrast, ``subjective'' speckle patterns form at the image plane of an optical system. Subjective patterns result from the interference of waves originating from different scattering regions within the lens. In this context, the random de-phasing of the waves adds to produce the speckle patterns \cite{goodman1976some,braga2003assessment}.

Speckle-based techniques have found widespread application across various fields, such as the monitoring of biological tissues, soft materials, and surface coatings. These methods offer a versatile tool for exploring physical, chemical, and biological processes by examining the temporal evolution of speckle patterns \cite{arizaga1999speckle,braga2003assessment}. Specifically, in this study, the DSPA method is employed to investigate the corrosion of the samples in different environments. By delving into the origins and nature of dynamic speckles, a deeper understanding of the internal dynamics during the corrosion process can be gained, facilitating more controlled experiments and simulations. Given the extensive information contained within DSPA, speckle-based methods are adaptable for studying a range of phenomena. These include detecting blood flow, monitoring scaffold activity, characterizing nanocomposites, tracking chemotaxis at low bacterial concentrations, assessing surface roughness and periodicity, detecting intralayer alignment in lipid layers, analyzing strawberry ripening, measuring tissue viscoelasticity, examining pitting corrosion, tracking parasite motility, investigating conduction mechanisms in memristor devices, evaluating seed health, and self-healing hydrogels \cite{braga2003assessment,fujii1985blood,farzam2020non,jamali2023surface,romero2009bio,panahi2022detection,mulone2013analysis,balamurugan2017study,pedram2023evaluation,abbasian2024dynamic, jamali2024speckle,pomarico2004speckle,sajjadi2024characterization}. Moreover, speckle patterns have been employed to control the motion of microparticles \cite{jamali2021speckle,sadri2024sorting,jamali2024speckletweezer,jamali2025specklerun}.

In this research, The aim is to monitor how AFC responds to chemical exposure in both acidic and alkaline environments and changes in its structure by using DSPA. To investigate these properties, we analyze various statistical parameters, including time history speckle pattern (THSP), co-occurrence matrix (COM), inertia moment (IM), Pearson correlation, roughness parameters like root mean square, skewness, and kurtosis.

This paper explores the application of DSPA to AFC. Section \ref{section2} provides a detailed explanation of the sample preparation and experimental procedure of dynamic speckle patterns. In section \ref{section3} the numerical processing background on the statistical analysis of DSPA is described. Section \ref{section4} presents the experimental findings and discusses the results in depth. Finally, the conclusions and implications of the study are summarized in Section \ref{section5}.

\section{Materials and methods}
\label{section2}
\subsection{Material preparation}
The tetraethoxysilane (28\%, TEOS), anhydrous ethanol (>99.7\%, EtOH), and n-hexane (>98\%) are purchased from Sigma-Aldrich company. N,N-Dimethylformamide (DMF, >99.5\%), and trimethylchlorosilane (TMCS, 98\%) are obtained from Merck company. Hydrochloric acid (HCl, >37\%) and ammonia (NH$_4$OH, >27\%) are sourced from Fisher Scientific. Portland cement (Grade P.O42.5), with a density of 3100 $\rm{kg/m^3}$ and a specific surface area of 300 $\rm{m^2/kg}$ as measured by the Blaine method, is used in the experiment. The YH-1 composite blowing agent, prepared using surfactants and various polymer components, is supplied by the Sika Group company.

Aerogel powder with an average particle size of 300 $\mu$m is used. The preparation of the aerogel powder involves three main stages: hydrolysis-condensation, aging modification, and ambient pressure drying. Initially, TEOS, H$_2$O, and HCl are added to the EtOH solvent and stirred at 40$^{\circ}$C for 2 hours. DMF, H$_2$O, and NH$_4$OH are then introduced sequentially, each followed by stirring for 5 minutes, leading to a polycondensation reaction at 40$^{\circ}$C for another 2 hours. After undergoing aging, surface modification, and drying at 80$^{\circ}$C for 24 hours, the aerogel powder is sieved to achieve the desired particle size. The final molar ratio of components (TEOS:EtOH:H$_2$O:DMF:HCl:NH$_4$OH) is 1:7:3:0.25:10$^{-5}$:3.57×10$^{-3}$. The resulting aerogel powder has a density of 170 $\rm{kg/m^3}$ and a porosity of 94\%. The preparation of aerogel cement composite is divided into three steps. First, cement, aerogel, and water are mixed to form a slurry based on a pre-set ratio. Due to the light and hydrophobic nature of silica aerogels, they tend to rise in the mixture, risking delamination of the foam. To address this, a thickener (0.15 wt\%) is added to coat the aerogel particles, improving consistency and viscosity. Additionally, a quick-setting admixture (3 wt\%) is introduced to ensure the slurry solidifies before aerogel flotation occurs, thereby preventing delamination.
Second, a mixture of foaming agents and water in a ratio of 1:40 is aerated using compressed air to produce a stable foam, which is then blended with the aerogel-cement slurry. The YH-1 composite foaming agent, known for its rapid foaming and stability, ensures the resulting foam remains fine and durable. Quick-setting agents are also incorporated to expedite solidification and minimize foam bursting during the process. Experimental observations indicate that P.O42.5 Portland cement does not significantly affect foam stability.
Finally, the aerogel cement slurry is molded and cured under controlled environmental conditions to achieve the desired material properties.

\subsection{Experimental procedure}

\begin{figure*}[t!]
	\begin{center}
		\includegraphics[width=0.9\linewidth]{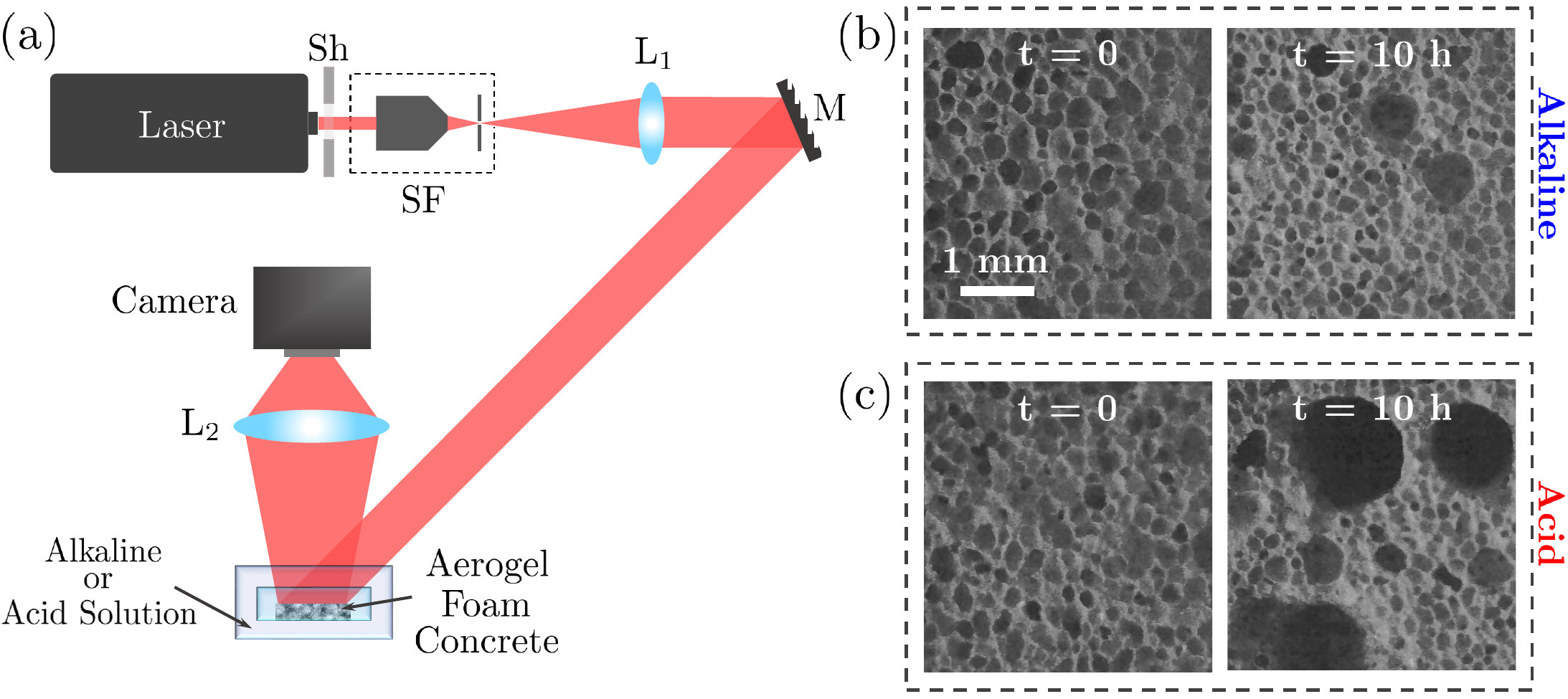}
		\caption{(a) Schematic of the experimental setup for DSPA in back-scattering mode; Sh, shutter, SF: spatial filter, L: lens, and M: mirror. (b) Time evolution of  AFC in an alkaline solution over a 10-hour period, showing visible changes. (c) Time evolution of AFC in an acidic solution over a 10-hour period, where the changes in the sample are more pronounced compared to the alkaline environment, indicating faster surface changes and increased destruction.}
		\label{Figure01}
	\end{center}
\end{figure*}

An optical speckle pattern is a granular interference pattern produced when coherent light, such as laser light, is scattered by a rough or inhomogeneous surface. When a coherent light beam interacts with the surface, the scattered waves interfere both constructively and destructively, resulting in a random distribution of bright and dark spots. This speckle pattern encodes valuable information about the physical and dynamic characteristics of the sample.

Changes in the sample, such as motion, biochemical activity, or other time-dependent processes, alter the speckle pattern over time. By analyzing these variations, important insights into the sample’s properties can be obtained.

The experimental setup for generating the dynamics speckle pattern is depicted in Fig.~\ref{Figure01}(a). A He-Ne laser (Red Dot Light, 632.8 nm, 2 mW, spot size at 100 mm is 1 mm) is directed through a spatial filter (SF) to remove unwanted spatial frequencies in the Fourier space using a pinhole positioned at the focal plane of a lens. The highly divergent emerging laser beam is collimated by using a lens (L$_1$) and then redirected by a mirror (M) onto the surface of the sample, which consists of AFC immersed in either acidic or alkaline solutions.

The backscattered light from the sample surface is collected by a second lens (L$_2$) and focused onto the sensor of a digital camera (Basler, acA800-200gm, 0.48 MP resolution, 10-bit dynamic range, 4.8 square pixel pitch). The interference of the backscattered light forms the speckle pattern, which is captured by the camera. The digital camera is set to record images at a resolution of 800$\times$600 pixels in the central area of the sensor, with appropriate exposure settings. The sequences of recorded speckle patterns are then transferred to a computer for further analysis. The He-Ne laser is chosen for its sufficient coherence and stability, making it ideal for speckle pattern analysis.

In addition to the experimental setup, Figure~\ref{Figure01}(b) and (c) present real micro-images of AFC samples in alkaline and acidic solutions, respectively, over a 10-hour period. 
Figure~\ref{Figure01}(b) shows the visual changes in the surface of the AFC in the alkaline solution at different time intervals. Over the course of 10 hours, the surface undergoes gradual alterations, with minimal visible surface degradation. These changes suggest a slower reaction and destruction process in the alkaline environment, as the surface maintains more of its initial texture during the experiment.
In contrast, Figure~\ref{Figure01}(c) depicts the AFC in the acidic solution, where the surface shows more significant changes over the same period. As early as 2 hours into the experiment, clear signs of degradation and corrosion are visible on the sample’s surface, with cracks and voids becoming more pronounced over time. The acidic environment accelerates the destruction process, leading to faster and more extensive surface damage compared to the alkaline solution. 

These images serve as visual evidence of the contrasting effects of acidic and alkaline environments on the AFC. The surface changes observed in these images will be quantitatively analyzed through roughness parameters using dynamic speckle pattern analysis (DSPA), which will be discussed in the subsequent sections.

In this experiment, the degradation and corrosion of cement materials are investigated using controlled acidic and basic environments to simulate real-world chemical weathering processes. A sulfuric acid (H$_2$SO$_4$) solution with a pH of 1.74 was used to replicate highly acidic conditions, akin to those caused by extreme acid rain, while a sodium hydroxide (NaOH) solution with a pH of 13.46 simulated a strongly alkaline environment. These pH values were chosen to represent the extremes of chemical exposure, emphasizing their relevance to environments where water-absorbing asphalts encounter both acidic precipitation and alkaline contaminants.

The acidic solution closely corresponds to conditions observed in acid rain, where pH levels typically range from 4.0 to 1.0 in heavily polluted areas. By simulating these conditions, the study provides critical insights into the impact of acid rain on infrastructure materials, particularly in regions subject to industrial emissions. The alkaline counterpart, although less common in natural precipitation, models scenarios of chemical exposure from industrial or construction environments, further broadening the scope of this work.

This controlled experimental approach allows for precise examination of surface degradation and roughness changes under chemically aggressive conditions. By enabling remote monitoring of material degradation, this method has significant implications for evaluating the long-term performance and durability of water-absorbing asphalts. Such insights are vital for designing resilient infrastructure capable of withstanding environmental challenges while maintaining functional efficiency.

\section{Numerical statistical processing}
\label{section3}
The speckle pattern arises from the interference of many waves with identical frequencies but varying phases and amplitudes. These waves add together simultaneously, leading to a resulting wave whose amplitude and consequently its intensity fluctuates randomly over time (spatiotemporal) due to the activity of the sample. The DSPA of active materials and dynamic samples allows for the evaluation of sample activity and provides important descriptions and information about their properties. The main aims of the presented research is to characterize the texture of the AFC in varied chemical environments. The numerical processing of recorded speckle patterns involves several useful statistical parameters, which are discussed in this section.

\subsection{Time history of speckle patterns (THSP)}

The time history of speckle patterns (THSP) is a two-dimensional matrix that captures the temporal evolution of sequential speckle patterns. To construct the THSP, M points are randomly selected from each of the  N consecutively acquired speckle patterns, and these  M$\times$ 1 columns are arranged side by side. As a result, the composite THSP image has dimensions 
M$\times$N, where the rows (M) correspond to the selected points, and the columns (N) reflect the intensity states of those points at each sampled time, representing both the temporal progression and the total number of patterns. This THSP matrix is used to analyze the activity in samples, with changes in intensity along the horizontal axis indicating activity. Larger intensity fluctuations along the horizontal lines of the THSP indicate higher activity levels in the sample \cite{braga2008time,BSLTLBOOK}.

\subsection{Co-occurrence matrix (COM)}

In addition to the graphical representation of activity provided by the THSP, it plays a crucial role in deriving various numerical metrics such as the co-occurrence matrix (COM) and others. The COM reflects the probability of intensity transitions between two adjacent pixels within the THSP. This intensity transition histogram characterizes the sample's activity by showing the spread of non-zero values away from the main diagonal.
The COM is mathematically defined as \cite{junior2016practical,BSLTLBOOK}:

\begin{equation}
	{\rm{COM}}(i,j)=\sum_{m=1}^{M}\sum_{n=1}^{N-1} 
	\begin{cases}
		1, & \text{if ~~ {\rm{THSP}}$(m,n) = i$} \\
		& \text{and~~{\rm{THSP}}$(m,n+1) = j$,}\\
		0, & \text{otherwise.}
	\end{cases}
	\label{eq:1}
\end{equation}
Here, $i$ and  $j$ represent the intensities of two adjacent pixels. For highly active samples, the intensity values evolve over time, causing an increase in non-zero elements near the main diagonal, giving the matrix a cloud-like distribution. In contrast, low-activity samples exhibit values that are more concentrated around the main diagonal \cite{arizaga1999speckle,BSLTLBOOK}.

\subsection{Inertia moment (IM)}
Inertia moment (IM) measures the dispersion of values around the principal diagonal of the co-occurrence matrix (COM).  IM is defined as the accumulated COM matrix values that are multiplied by the square distance to the original diagonal and normalized \cite{junior2016practical,zdunek2014biospeckle,BSLTLBOOK}:
\begin{equation}
	{\rm{IM}} = \sum_{i}^{}\sum_{j}^{} \frac{{\rm{COM}}(i,j)}{\sum_{m}{\rm{COM}}(i,m)}|i-j|^2.
	\label{eq:2}
\end{equation}
Normalization reduces the impact of image heterogeneity by ensuring that the sum of occurrence values in each row of the COM equals 1 \cite{arizaga1999speckle,BSLTLBOOK}. IM provides a valuable statistical measure for assessing sample activity, with higher IM values indicating increased destruction in AFC.

\subsection{Pearson correlation coefficient of speckle patterns}
Pearson correlation coefficient can quantify the similarity between speckle patterns at different times, providing insight into the underlying dynamics of the sample. The Pearson correlation coefficient, \( C_{Pr} \), measures the linear relationship between two variables. In the context of DSPA, it helps to quantify the temporal correlation between speckle patterns at two different time points, \( t_1 \) and \( t_2 \). A high Pearson correlation indicates that the speckle patterns are similar, implying slow or minimal changes in the sample, while a low correlation indicates rapid changes or significant dynamics.

For two speckle patterns at times \( t_1 \) and \( t_2 \), let the intensities at a specific pixel \( i \) be denoted as \( I_{t_1i} \) and \( I_{t_2i} \), respectively. The Pearson correlation coefficient between these two-time points is defined as \cite{cohen2009pearson,BSLTLBOOK}:

\begin{equation}
    C_{Pr_{t_1 t_2}} = \frac{\sum_{i=1}^{n} \langle(I_{t_1i} - \langle {I}_{t_1} \rangle)(I_{t_2i} - \langle {I}_{t_2} \rangle)\rangle}{\sqrt{\sum_{i=1}^{n} (I_{t_1i} - \langle {I}_{t_1} \rangle)^2} \sqrt{\sum_{i=1}^{n} (I_{t_2i} - \langle {I}_{t_2} \rangle)^2}},
    \label{eq:3}
\end{equation}
where:
\begin{itemize}
    \item \( I_{t_1i} \) and \( I_{t_2i} \) represent the intensities at pixel \( i \) at times \( t_1 \) and \( t_2 \), respectively,
    \item \( \langle {I}_{t_1} \rangle \) and \( \langle {I}_{t_2} \rangle \) are the mean intensity values of the speckle patterns at times \( t_1 \) and \( t_2 \),
    \item \( n \) is the total number of pixels in the speckle pattern.
\end{itemize}

By calculating the Pearson correlation for a sequence of time points in the THSP, we can evaluate the degree of similarity between speckle patterns as the sample evolves. This can be used to infer the dynamics within the sample \cite{sedgwick2012pearson,BSLTLBOOK}.

\begin{figure*}[t!]
	\begin{center}
		\includegraphics[width=0.9\linewidth]{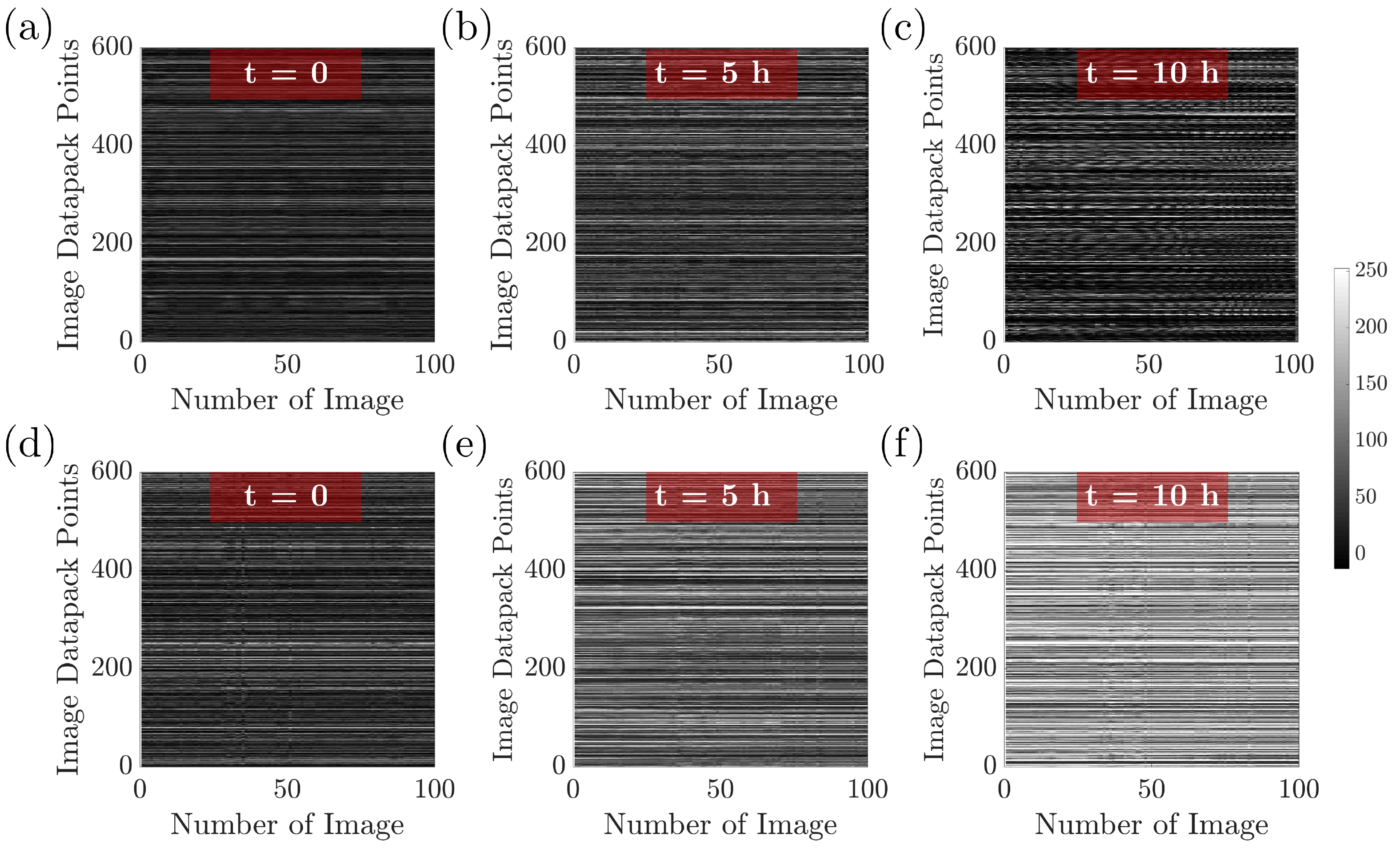}
		\caption{Time history of speckle patterns (THSP) analysis for AFC in acidic and alkaline environments over time, THSP matrices are formed by tracking 600 random points throughout a collection of 100 speckle patterns. (a-c) THSP matrices for AFC in an alkaline solution at initial, 5 hours, and 10 hours, respectively. (d-f) THSP matrices for AFC in an acidic solution at corresponding time intervals. The brightness and dynamic changes in the THSP reveal greater activity and more rapid destruction in the acidic environment compared to the alkaline solution, as indicated by more pronounced intensity fluctuations and structural changes over time.}

		\label{Figure02}
	\end{center}
\end{figure*}

\subsection{Roughness analysis}
By considering the speckle pattern intensity matrix as a two-dimensional statistical distribution, it is possible to define roughness parameters that can be used independently of THSP for speckle pattern-based evaluations. It is particularly notable that when speckle patterns are generated through the scattering of laser light from a rough surface (such as AFC), the roughness of the speckle pattern matrix is directly related to the roughness of the sample \cite{jeyapoovan2012statistical}. This correlation between intensity variations and surface roughness presents a valuable tool for characterizing surfaces and roughness parameters. However, the roughness of the speckle patterns can also serve as an independent measure for assessing the internal activities of samples, irrespective of the mechanism leading to the formation of the speckle patterns 
 \cite{jamali2023surface,pedram2023evaluation}.

A comprehensive and reliable analysis of the intensity distribution can be achieved by using the set of roughness parameters, which include average roughness (R$_{\rm{av}}$), root mean square roughness (R$_{\rm{rms}}$), skewness (R$_{\rm{sk}}$), and kurtosis (R$_{\rm{ku}}$). These parameters correspond to the first, second, third, and fourth moments of the deviation of the data from the mean of the distribution, respectively.

The average roughness (R$_{\rm{av}}$) quantifies the average deviation of intensity values from their mean across the entire matrix:
\begin{equation}
    {\rm{R_{\rm{av}}}} = \frac{1}{P~Q} \sum_{p=1}^{P} \sum_{q=1}^{Q}|I(p,q) - \langle{I}(p,q)\rangle|.
    \label{eq:4}
\end{equation}
This metric provides insight into the overall variation of pixel intensities within the speckle patterns. Meanwhile, the root mean square roughness (R$_{\rm{rms}}$) indicates the standard deviation of the intensity distribution:
\begin{equation}
    {\rm{R_{\rm{rms}}}} = \bigg[\frac{1}{P~Q} \sum_{p=1}^{P} \sum_{q=1}^{Q}\left[I(p,q) - \langle{I}(p,q)\rangle\right]^2\bigg]^{\frac{1}{2}}, 
    \label{eq:5}
\end{equation}
where $P$ and $Q$ represent the horizontal and vertical dimensions of the speckle pattern matrix, $p$ and $q$ are pixel indices, and $I$ denotes the intensity value at each pixel. This parameter provides a measure of the spread of intensity values around the mean, reflecting the variability in surface roughness.

In addition to R$_{\rm{av}}$ and R$_{\rm{rms}}$, skewness (R$_{\rm{sk}}$) and kurtosis (R$_{\rm{ku}}$) offer additional descriptive insights into the intensity distribution:
\begin{equation}
    {\rm{R_{\rm{sk}}}} = \frac{1}{P~Q~S_2^3} \sum_{p=1}^{P} \sum_{q=1}^{Q}\left[I(p,q) - \langle{I}(p,q)\rangle\right]^3,
    \label{eq:6}
\end{equation}
\begin{equation}
    {\rm{R_{\rm{ku}}}} = \frac{1}{P~Q~S_2^4} \sum_{p=1}^{P} \sum_{q=1}^{Q}\left[I(p,q) - \langle{I}(p,q)\rangle\right]^4.
    \label{eq:7}
\end{equation}
R$_{\rm{sk}}$, or skewness, measures the symmetry of the intensity distribution. For a perfectly symmetrical distribution, skewness is zero, as the deviations from the mean in both directions are balanced. A negative skewness value suggests a predominance of lower intensities (valleys), while a positive skewness indicates a distribution dominated by higher intensities (peaks).
Kurtosis (R$_{\rm{ku}}$) evaluates the sharpness or flatness of the distribution. For a normal distribution, kurtosis is typically 3. A kurtosis greater than 3 implies that the distribution has sharper peaks or deeper valleys, while a kurtosis less than 3 suggests a broader, flatter distribution, often corresponding to more gradual intensity variations within the speckle patterns \cite{jeyapoovan2012statistical, gadelmawla2002roughness}.

\section{Result and discussion}
\label{section4}

\begin{figure*}[t!]
	\begin{center}
		\includegraphics[width=0.9\linewidth]{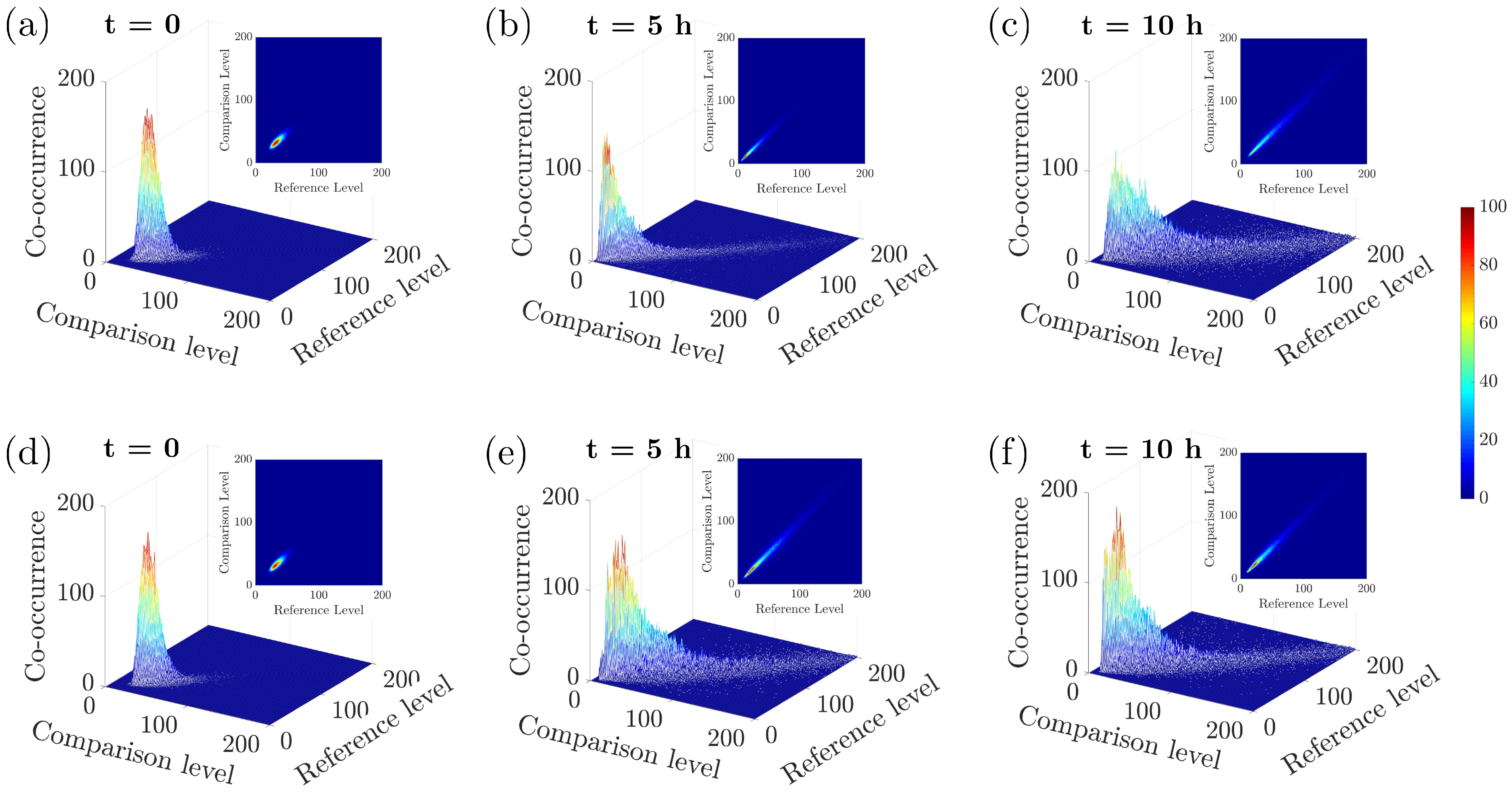}
		\caption{Co-occurrence matrix (COM) analysis of AFC destruction process in acidic and alkaline solutions. (a-c) 3D graphs and 2D maps of COM matrices for AFC in alkaline solution at initial, 5 hours, and 10 hours, respectively. (d-f) COM matrices for AFC in acidic solution at corresponding time intervals. The dispersion of COM values along the principal diagonal indicates surface texture changes, with greater spreading in the acidic environment, suggesting a more rapid destruction process compared to the alkaline solution.}

		\label{Figure03}
	\end{center}
\end{figure*}

\begin{table}[ht]
\centering
\resizebox{1\columnwidth}{!}{%
\begin{tabular}{@{}lccccc@{}}
\toprule
\textbf{Sample} & \textbf{Initial} & \textbf{Final in} & \textbf{Weight Loss} & \textbf{Final in} & \textbf{Weight Loss} \\
\textbf{ID} & \textbf{weight (g)} & \textbf{H$_2$SO$_4$ (g)} & \textbf{in H$_2$SO$_4$ (\%)} & \textbf{NaOH (g)} & \textbf{in NaOH (\%)} \\ \midrule
1            & 5.12            & 4.16              & 9.60                 & 4.92            & 2.00                \\
2            & 5.09            & 3.86            & 12.30                  & 4.86            & 2.30                \\
3            & 6.53             & 5.49             & 10.40                 & 6.21           & 3.20                 \\
4            & 4.60             & 3.42             & 11.80                 & 4.35            & 2.50                \\
5            & 4.89             & 3.73             & 11.60                & 4.67           & 2.20                \\ \bottomrule
\end{tabular}
}
\caption{Weight measurements of AFC before and after immersion in acidic (H$_2$SO$_4$, pH=1.74) and alkaline (NaOH, pH=13.46) solutions, showing weight loss percentages for each environment.}
\label{table1}
\end{table}

The THSP images provide valuable analysis for monitoring the temporal evolution of speckle patterns generated by the surface of AFC during chemical reactions in acidic and alkaline environments. This technique is critical for tracking changes over time and capturing the dynamic activity and durability of the sample. As the chemical reaction progresses, the speckle pattern evolves in both distribution and intensity, reflecting alterations in the surface texture of the AFC. Figure~\ref{Figure02} presents the THSP matrices for AFC immersed in both acidic and alkaline solutions at different time intervals.
In Fig.~\ref{Figure02}, the THSP matrices for AFC are illustrated for both environments. Subfigures \ref{Figure02}(a-c) correspond to the THSP in alkaline solution at the initial time, after 5 hours, and after 10 hours, respectively. Similarly, subfigures \ref{Figure02}(d-f) depict the THSP in an acidic solution at the same time intervals. These THSPs are generated by compiling the intensity data of 600 randomly selected pixels from a series of 100 speckle patterns. Significant intensity fluctuations at these points indicate internal dynamic processes within the AFC structure.
Over the course of 10 hours, the appearance and disappearance of distinct bright horizontal lines, along with the formation of discontinuous patterns within the THSP, suggest progressive changes in the material. High activity levels are marked by the transformation of the THSP into a pattern resembling a typical speckle field, where the bright lines become less distinct, indicating an increasingly random structure.
Notably, the THSP matrix for the acidic environment appears significantly brighter than that for the alkaline solution, particularly as the reaction progresses. This increased brightness in the acidic THSP indicates more pronounced activity and a faster degradation process in acidic conditions compared to the alkaline environment. These results clearly demonstrate that AFC undergoes more rapid deterioration in an acidic solution, as evidenced by the more intense dynamic changes captured in the THSP.

Building on the insights gained from THSP analysis, a further exploration of the destruction process of AFC can be achieved through the analysis of the co-occurrence matrix (COM). COM offers a deeper understanding of the microstructural changes by quantifying the spatial relationships between pixel intensities within the THSP over time.
COM captures the spatial relationships between pixel intensities in the THSP, which is crucial for assessing the textural evolution of AFC as it undergoes chemical reactions and destruction. As the concrete deteriorates, the surface texture becomes more uniform, creating more voids and openings, which is reflected in the COM. This matrix provides critical insights into the microstructural evolution during the destruction process. To better understand the variations in activity over time, we calculated the corresponding COM matrices at different intervals. 
Figure~\ref{Figure03}(a-c) shows the 3D plots and 2D projections of the COM matrix for AFC in an alkaline solution, illustrating its temporal evolution over a 10-hour period. In contrast, Figure~\ref{Figure03}(d-f) presents the COM analysis for the sample in an acidic environment under the same conditions.

The reference intensity levels, \(i\) and \(j\), used in the COM matrix calculations, represent the intensity transitions between neighboring pixels, as defined in Eq.~\ref{eq:1}.

Two key observations emerge from the COM analysis in Fig.~\ref{Figure03}: (1) As time progresses, the points along the principal diagonal of the COM matrix gradually disperse, forming a cloud-like structure, and (2) the number of points with very high COM values decreases as the experiment continues. According to the definitions of COM and THSP, higher sample activity corresponds to more frequent and pronounced intensity fluctuations over time. Thus, the clustering of values around the principal diagonal suggests a more homogeneous texture, while the presence of non-zero elements farther from the diagonal indicates stronger fluctuations within the sample.

The COM matrix is typically normalized, resulting in what is referred to as the "modified co-occurrence matrix," which represents the transition probabilities between different intensity levels in the THSP. While the total number of high COM values may not offer additional insight into the sample, the analysis clearly indicates that AFC in acidic solution exhibits greater activity and a more rapid destruction process compared to that in alkaline solution. This is evidenced by the more extensive dispersion of COM values and stronger intensity fluctuations in the acidic environment.

Table \ref{table1} shows the weight loss measurements of AFC samples after immersion in acidic (H$_2$SO$_4$) and alkaline (NaOH) solutions, illustrating the degradation caused by these chemical environments. The weight loss observed in AFC samples after exposure to acidic and alkaline environments highlights the degradation caused by these chemical conditions. Samples exposed to the sulfuric acid (H$_2$SO$_4$) solution (pH=1.74) exhibited a more significant weight reduction, with an average weight loss of approximately 6\%, compared to an average weight loss of about 2.5\% in the NaOH solution (pH=13.46). This discrepancy underscores the destructive impact of acid rain, which is simulated here by the acidic solution.
Acid rain, characterized by pH values similar to those of the H$_2$SO$_4$ solution used in this study, poses severe environmental hazards, particularly to infrastructure made of cementitious materials. Over time, such acidic exposure can lead to structural weakening, surface erosion, and increased maintenance costs.

Our method, utilizing optical techniques for remote monitoring of material degradation, offers a practical solution to detect early signs of damage caused by acid rain. This capability is crucial for proactive infrastructure management, enabling timely interventions to mitigate long-term environmental and economic consequences.

\begin{figure*}[t!]
	\begin{center}
		\includegraphics[width=0.9\linewidth]{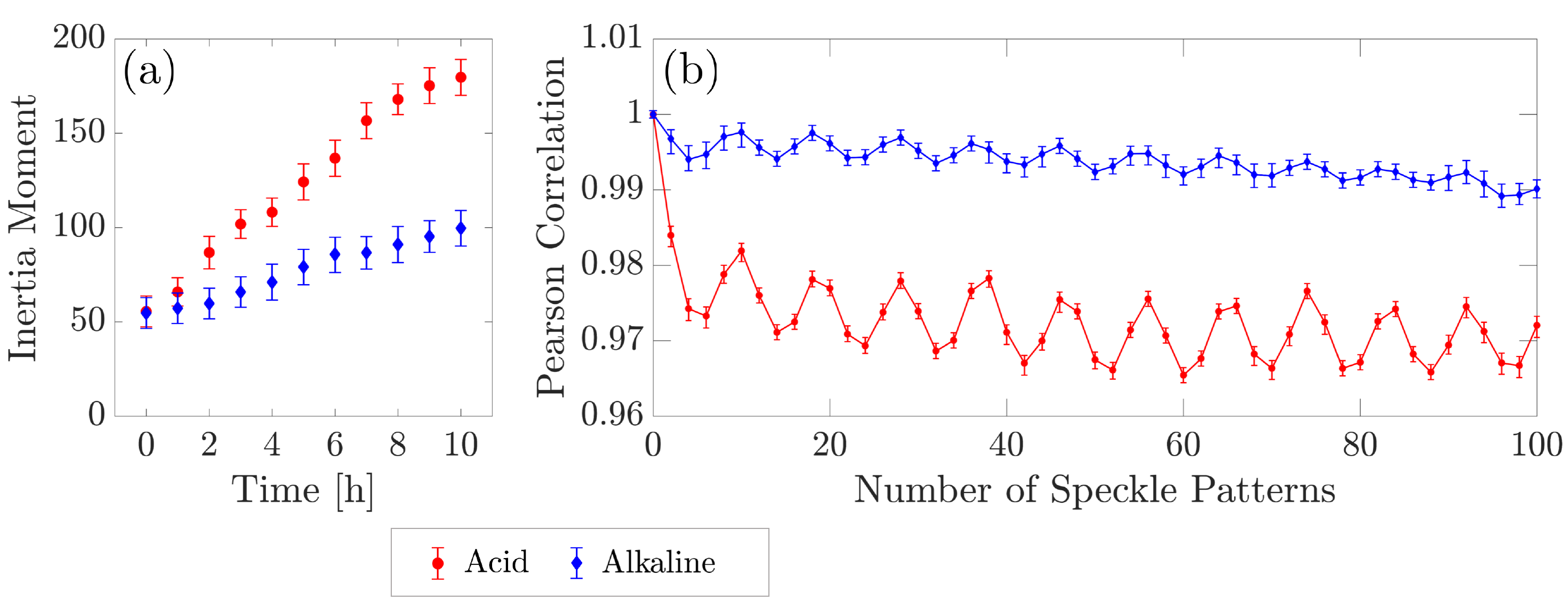}
		\caption{Inertia Moment (IM) and Pearson correlation analysis for AFC in acidic and alkaline environments. (a) Average IM values over 10 hours, measured every 2 hours. Higher IM values in the acidic solution indicate greater heterogeneity and more rapid destruction. (b) Pearson correlation coefficients for speckle patterns, showing greater correlation (i.e., stability) in the alkaline environment and more rapid changes in the acidic environment, reflecting faster destruction and corrosion. Error bars represent standard deviations across five samples.}

		\label{Figure04}
	\end{center}
\end{figure*}

Following the COM analysis, further quantitative evaluation of the destruction and corrosion process in AFC is carried out by calculating the IM parameter and the Pearson correlation of speckle patterns. These parameters offer additional insight into the internal activity of the samples, complementing the information provided by the THSP and COM.

The IM quantifies the spread of COM values around the principal diagonal and serves as an indicator of the degree of destruction and corrosion in the material over time. Additionally, Pearson correlation analysis is performed to evaluate the similarity of the speckle patterns at different time points, providing a statistical measure of how the speckle pattern evolves under different conditions.
IM is a metric that assesses the distribution of speckle intensities relative to the mean intensity, providing a measure of the pattern’s heterogeneity. A higher IM value indicates a more heterogeneous speckle pattern, which is characteristic of a sample undergoing significant destruction. Figure~\ref{Figure04}(a-b) presents two key statistical parameters: the IM and the Pearson correlation coefficient of the speckle patterns. These parameters are derived from at least five measurements for each condition, and the corresponding averages and standard deviations are reported.
Figure~\ref{Figure04}(a) illustrates the evolution of the average IM values over time for AFC in both acidic and alkaline solutions, with data points recorded every two hours. The IM is calculated by summing the squared row distances from the principal diagonal of the THSP, making it an effective quantitative measure of the ``cloudiness'' or dispersion of the COM distribution around its diagonal. As shown in the figure, the IM values increase over time, signaling the ongoing destruction and corrosion of the AFC samples throughout the experiment. The error bars in Figure~\ref{Figure04}(a) represent the standard deviations across five different measurements for each sample type.
The increasing IM values for both environments can be attributed to heightened activity during the internal chemical interactions within the material. However, it is evident that the samples in the acidic solution exhibit significantly higher IM values compared to those in the alkaline solution. This result suggests that the destruction and corrosion process occurs more rapidly in the acidic environment, consistent with the findings from the THSP and COM analyses.

In addition to the IM analysis, Pearson correlation is used to assess the similarity between speckle patterns at different time intervals. Figure~\ref{Figure04}(b) shows the Pearson correlation coefficients, which quantify the degree of correlation between successive speckle patterns. Higher Pearson correlation values indicate that the speckle patterns are more consistent over time, suggesting less dramatic internal changes. In the alkaline environment, the speckle patterns maintain a higher degree of correlation compared to the acidic environment, where the rapid destruction process leads to greater divergence between speckle patterns. This highlights that while the AFC in acidic conditions undergoes more intense and rapid changes, the samples in the alkaline solution experience a more gradual and uniform evolution.

Following the evaluation of inertia moment and Pearson correlation, further analysis of the surface characteristics of AFC samples is conducted using statistical roughness parameters. These parameters, including root mean square (RMS), skewness, and kurtosis, provide additional insight into the ongoing destruction and corrosion process by characterizing the intensity distribution of the related speckle patterns.

\begin{figure*}[t!]
	\begin{center}
		\includegraphics[width=0.9\linewidth]{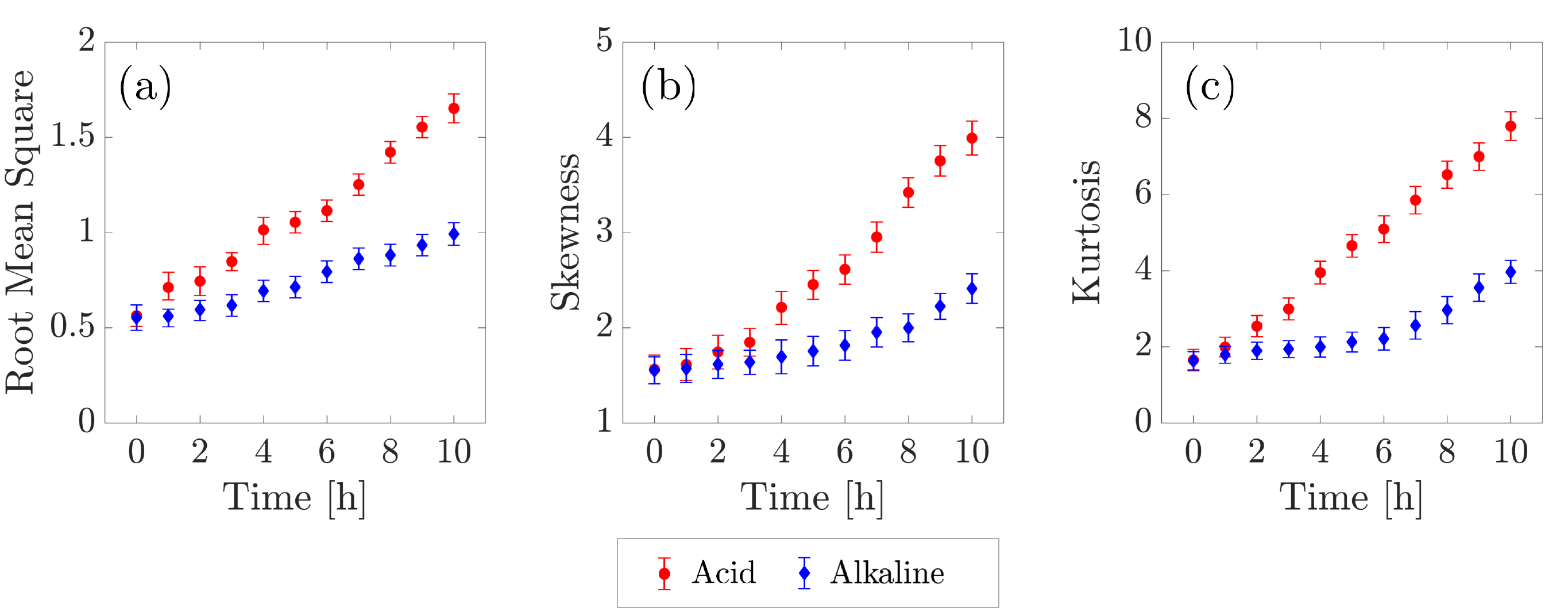}
		\caption{Analysis of root mean square (RMS), Skewness, and Kurtosis for AFC in acidic and alkaline environments. (a) RMS values indicate increasing surface roughness, with a sharp rise in the acidic solution after 2 hours, reflecting faster destruction. (b) skewness values, show increasing asymmetry in intensity distribution, especially in the acidic environment after 2 hours. (c) Kurtosis values indicate sharper intensity distribution peaks, with a more rapid rise in the acidic environment, suggesting more extreme surface changes. Error bars represent the standard deviations from five samples for each measurement.}
		\label{Figure05}
	\end{center}
\end{figure*}

In addition to IM, the roughness parameters offer crucial information about the surface structure of the samples. For porous materials like AFC, the roughness characteristics are particularly significant in reflecting the surface changes caused by chemical interactions. The differences in surface structure observed in both acidic and alkaline conditions can be attributed to the increased interaction between the sample and the corrosive environment, leading to variations in the corresponding statistical roughness parameters.

Figure~\ref{Figure05}(a) illustrates the evolution of the RMS values over time, capturing the progressive destruction of the samples. Initially, the RMS values for both acidic and alkaline solutions follow a similar trend. However, a sharp increase in the RMS value is observed for the acidic environment after 2 hours, and by 4 hours, the two curves diverge significantly. This rise in RMS for the acidic solution indicates a more rapid degradation process, suggesting a higher degree of destruction compared to the alkaline condition. 

The RMS parameter is sensitive to changes in surface or bulk properties, such as the formation and widening of cracks and voids due to destruction and corrosion. This increase in RMS reflects the growing heterogeneity of the material’s structure as the destruction progresses, particularly in the acidic environment.

In addition to RMS, skewness, and kurtosis provide further insight into the distribution of speckle pattern intensities, which are linked to the surface roughness and asymmetry of the sample. Skewness measures the asymmetry of the intensity distribution, while kurtosis quantifies the peakedness or sharpness of the distribution. Figure~\ref{Figure05}(b) shows the variation in the skewness parameter over time, where each data point represents the average skewness value calculated from 100 speckle patterns per sample in both environments, taken every two hours.

The observed increase in skewness after 2 hours, particularly in the acidic solution, suggests that the intensity distribution becomes more asymmetric, with extreme values becoming more prominent. This trend indicates that the surface roughness increases as cracks and holes form during the destruction and corrosion process, leading to a more uneven intensity distribution in the speckle pattern. The acidic environment shows a more pronounced skewness rise, indicating more aggressive surface changes compared to the alkaline solution.

Figure~\ref{Figure05}(c) presents the kurtosis values, which describe the sharpness of the intensity distribution peaks. An increase in kurtosis signifies the presence of very sharp intensity points in the speckle field, corresponding to extreme scattering events caused by the ongoing destruction. According to the results shown in Fig.~\ref{Figure05}(c), the kurtosis trend exhibits a steady increase over time, with the acidic solution showing a more rapid rise, reflecting more extreme surface changes and scattering events in comparison to the alkaline solution.

In Figures ~\ref{Figure05}(a-c), the error bars represent the standard deviation calculated from five samples for each measurement, providing insight into the consistency of the roughness parameters across different samples. The results clearly demonstrate significant differences in surface roughness and intensity distribution between the acidic and alkaline environments, with the acidic condition leading to more pronounced non-uniformities and more rapid progression of destruction and corrosion.

This study demonstrates a novel optical approach for evaluating the degradation of cement materials under sulfuric acid (H$_2$SO$_4$) and alkaline (NaOH) conditions. By simulating the effects of acid rain, the research provides valuable insights into the durability and efficiency of water-absorbing asphalts. The ability to remotely monitor changes in surface roughness highlights the practical application of this method in assessing material performance in real-world environments. This non-invasive technique offers a promising tool for evaluating infrastructure sustainability under various environmental stresses.

\section{Conclusion}
\label{section5}
In conclusion, we systematically investigated remote characterization of the destruction and corrosion processes of AFC in acidic and alkaline environments using speckle pattern analysis and associated roughness parameters. The experimental procedure utilized a He-Ne laser setup to generate optical speckle patterns, which were captured and analyzed through various metrics such as the time history of speckle patterns (THSP), co-occurrence matrix (COM), inertia moment (IM), Pearson correlation, and statistical roughness parameters (root mean square, skewness, and kurtosis). These metrics provided critical insights into the evolution of surface changes and internal activities during the chemical degradation of AFC.
The THSP analysis revealed that AFC in the acidic environment showed more pronounced activity and faster destruction compared to the alkaline solution, with greater intensity fluctuations and structural changes. COM analysis confirmed this, showing a wider dispersion of intensity transitions and greater heterogeneity in the acidic condition. The increasing IM values correlated with faster deterioration in the acidic environment, while the Pearson correlation indicated more gradual changes in the alkaline solution.
Additionally, the roughness parameters—RMS, skewness, and kurtosis—demonstrated more aggressive surface degradation in the acidic solution. The RMS rose sharply after 2 hours, indicating significant surface roughness due to cracks and voids, while skewness and kurtosis increases further highlighted the more severe destruction in acidic conditions
Overall, this study provides a detailed characterization of the destruction process in AFC, showing that acidic environments lead to more rapid and extensive corrosion compared to alkaline conditions. The combination of speckle pattern analysis and roughness parameter evaluation proves to be a robust method for monitoring and quantifying surface and internal changes in such materials. These findings and methodology offer valuable insights for future applications in material science. Also, remote characterization of these samples particularly for the development of more resilient materials that can withstand harsh chemical environments.

\section*{Author contributions statement}
R.J. and M.H.S.  carried out the speckle experiments. R.J. analyzed and interpreted the data.  R.j. conceived the project.
All authors discussed the results and contributed to the writing and reviewing of the manuscript. A.R.M. supervised the project. 

\section*{Competing interests}
The authors declare no competing interests.
\vspace{-0.1 cm}
\section*{Data availability}
The datasets used or analyzed during the current study available from the corresponding author on reasonable request.

\nocite{*}

\bibliographystyle{unsrt} 

\end{document}